%% file: airborne_virus_spread_statistical_transmission_model.tex
\documentclass[sigconf]{acmart}

\renewcommand\footnotetextcopyrightpermission[1]{}
\usepackage{hyperref}

\usepackage{acronym}
\usepackage{amsmath}
\usepackage[english]{babel}
\usepackage{bbm}
\usepackage{bm}
\usepackage{caption}
\usepackage{cleveref}
\usepackage{enumerate}
\usepackage[T1]{fontenc}
\usepackage{graphicx}
\usepackage[utf8]{inputenc}
\usepackage{siunitx}

\newcommand{\dtau}{\ensuremath{\textrm{d}\tau}}
\newcommand{\ttx}{\ensuremath{t^{\textrm{tx}}}}
\newcommand{\trx}{\ensuremath{t^{\textrm{rx}}}}
\newcommand{\uex}{\ensuremath{u^{\textrm{ex}}}}
\newcommand{\uin}{\ensuremath{u^{\textrm{in}}}}
\newcommand{\Fex}{\ensuremath{F^{\textrm{ex}}}}
\newcommand{\Fin}{\ensuremath{F^{\textrm{in}}}}
\newcommand{\Htx}{\ensuremath{H^{\textrm{tx}}}}
\newcommand{\Btx}{\ensuremath{B^{\textrm{tx}}}}
\newcommand{\Ltx}{\ensuremath{L^{\textrm{tx}}}}
\newcommand{\Ctx}{\ensuremath{C^{\textrm{tx}}}}
\newcommand{\Stx}{\ensuremath{S^{\textrm{tx}}}}
\newcommand{\Hrx}{\ensuremath{H^{\textrm{rx}}}}
\newcommand{\Brx}{\ensuremath{B^{\textrm{rx}}}}
\newcommand{\Lrx}{\ensuremath{L^{\textrm{rx}}}}
\newcommand{\Crx}{\ensuremath{C^{\textrm{rx}}}}
\newcommand{\Srx}{\ensuremath{S^{\textrm{rx}}}}
\newcommand{\FEex}{\ensuremath{\textrm{FE}^{\textrm{ex}}}}
\newcommand{\FEin}{\ensuremath{\textrm{FE}^{\textrm{in}}}}
\newcommand{\MFex}{\ensuremath{\textrm{MF}^{\textrm{ex}}}}
\newcommand{\GFex}{\ensuremath{\textrm{GF}^{\textrm{ex}}}}
\newcommand{\MFin}{\ensuremath{\textrm{MF}^{\textrm{in}}}}
\newcommand{\GFin}{\ensuremath{\textrm{GF}^{\textrm{in}}}}
\newcommand{\EIA}{\ensuremath{\textrm{EIA}}}
\newcommand{\IIA}{\ensuremath{\textrm{IIA}}}
\newcommand{\Sch}{\ensuremath{S^\textrm{ch}}}
\newcommand{\Tin}{\ensuremath{T^\textrm{in}}}

\acrodef{MC}{Molecular Communications}
\acrodef{sarscovtwo}[SARS-CoV-2]{severe acute respiratory syndrome coronavirus 2}
\acrodef{COVID-19}{coronavirus disease 2019}
\acrodef{IA}{infectious aerosol}
\acrodef{IIA}{inhaled infectious aerosol}
\acrodef{IP}{infected person}
\acrodef{HP}{healthy person}
\acrodef{RV}{random variable}

\setcopyright{acmcopyright}
\copyrightyear{}
\acmYear{}
\acmDOI{}

\acmConference[NANOCOM '21]{NANOCOM '21}{September 07--09, 2021}{Virtual Conference}

\begin{document}
\title{Statistical Modeling of Airborne Virus Transmission Through Imperfectly Fitted Face Masks}

\author{Sebastian Lotter, Lukas Brand, Maximilian Schäfer, and Robert Schober}
\affiliation{%
    \institution{Friedrich-Alexander University, Erlangen-Nuremberg, Germany}
    \country{}
}
\email{{sebastian.g.lotter,lukas.brand,max.schaefer,robert.schober}@fau.de}

\begin{abstract}
\vspace*{-0.5ex}
The rapid emergence and the disastrous impact of the \ac{sarscovtwo} pandemic on public health, societies, and economies around the world has created an urgent need for understanding the pathways critical for virus transmission and counteracting the spread of \ac{sarscovtwo} efficiently.
Airborne virus transmission by asymptomatic \ac{sarscovtwo}-infected individuals is considered to be a major contributor to the fast spread of \ac{sarscovtwo} and social distancing and wearing of face masks in public have been implemented as countermeasures in many countries.
Concerted research efforts in diverse scientific fields have meanwhile advanced the understanding of the physical principles of the manifold processes involved in airborne transmission of \ac{sarscovtwo}.
As part of these efforts, the physics and dynamics of aerosol filtration by various types of face masks have been studied.
However, a comprehensive risk assessment framework for the airborne transmission of \ac{sarscovtwo} incorporating realistic assumptions on the filtration of \acp{IA} by face masks is not available yet. 
In particular, in most end-to-end models for airborne virus transmission, it is neglected that the stochastic spread of \acp{IA} through imperfectly fitted face masks depends on the dynamics of the breathing of the wearer.
In this paper, we consider airborne virus transmission from an infected but asymptomatic individual to a healthy individual, both wearing imperfectly fitted face masks, in an indoor environment.
By framing the end-to-end virus transmission as a \ac{MC} system, we obtain a statistical description of the number of \acp{IA} inhaled by the healthy person subject to the respective configurations of the face masks of both persons.
We demonstrate that the exhalation and inhalation air flow dynamics have a significant impact on the stochastic filtering of \acp{IA} by the imperfectly fitted face masks.
Furthermore, we are able to show that the fit of the face mask of the infected person can highly impact the infection probability if the infectious dose for virus transmission to the healthy person is in a critical range.
We conclude that the proposed \ac{MC} model may contribute a valuable assessment tool to fight the spread of \ac{sarscovtwo} as it naturally encompasses the randomness of the transmission process and thus enables comprehensive risk analysis beyond statistical averages.
\end{abstract}

\maketitle
\pagestyle{plain}

\acresetall
\section{Introduction}
\input{introduction}

\section{Physical Transmission Scenario}\label{sec:physical_transmission_scenario}
\input{physical_transmission_scenario}

\section{Transmitter}\label{sec:transmitter}
\input{transmitter}

\section{Channel}\label{sec:channel}
\input{channel}

\section{Receiver}\label{sec:receiver}
\input{receiver}


\section{Numerical Results}\label{sec:numerical_results}
\input{numerical_results}

\section{Conclusions}\label{sec:conclusions}
\input{conclusions}

\bibliographystyle{ACM-Reference-Format}
\bibliography{gc21_etal}

\end{document}

%% file: introduction.tex
\vspace*{-0.5ex}
The global outbreak of the {\em \ac{COVID-19}} \cite{wiersinga20} since fall 2019 has impacted and is impacting societies all over the world causing large-scale and long-term damage to public health, society, and economy.
\ac{COVID-19} is caused by a novel type of coronavirus, {\em \ac{sarscovtwo}}, which is highly transmissible between humans \cite{hu21}.
To contain the spread of \ac{sarscovtwo}, it has soon been advocated to target the airborne transmission of \ac{sarscovtwo} in indoor environments, since this transmission route is expected to contribute significantly to the spread of the virus \cite{morawska20,azimi21}.
Airborne viral transmission in {\em poorly ventilated} indoor environments by {\em asymptomatic} individuals is especially relevant, because, in the absence of proper ventilation, the \acp{IA}, i.e., the virus carrying microdroplets, possibly remain suspended in the room air for a long time and asymptomatic individuals have no indication to quarantine themselves \cite{lelieveld20}.
To minimize the risk of airborne transmission of \ac{sarscovtwo}, face masks have been widely and successfully used in practice \cite{gandhi21}.
At the same time, experts with diverse professional backgrounds have studied the impact of face masks on the airborne transmission of \ac{sarscovtwo} experimentally and in modeling studies \cite{howard21}.
While these efforts have advanced the scientific understanding of many processes and parameters relevant for airborne virus transmission, a comprehensive assessment of the impact of face masks on the infection risk remains challenging.
Specifically, the end-to-end modeling of airborne \ac{sarscovtwo} transmission starting with the exhalation of \acp{IA} by an \ac{IP} and ending with the inhalation of \acp{IA} by a \ac{HP} is complicated by the inherent randomness of the process.
This randomness originates from the small number of \acp{IA} typically involved in virus transmission \cite{cheng20} and it implies that a statistical characterization of the transmission (beyond the mean) is required to fully assess the infection risk.
However, existing transmission models tend to neglect this randomness in favor of a comprehensible mathematical model and communicable outcomes \cite{mittal20a}.
Additional randomness arises if the face mask is not fitted perfectly to the face of the wearer.
In this case, part of the breathing airflow and, consequently, some of the exhaled \acp{IA} do not pass the mask filter piece but leak through gaps at the edge of the mask.
This leakage flow has been attributed significance for airborne virus transmission \cite{viola21} and needs therefore to be accounted for.
However, in existing end-to-end models for airborne virus transmission, it is often not taken into account.
In order to overcome these limitations, in this paper, we explore the modeling of airborne virus transmission under the framework of \ac{MC}, an interdisciplinary research field that has recently emerged at the intersection of information and communication theory and biology \cite{nakano13}.

The \ac{MC} paradigm has been applied to model airborne virus transmission in \cite{khalid19,khalid20,gulec20,gulec21,schurwanz20}.
However, the impact of face masks on the statistics of the end-to-end airborne virus transmission has not been considered.
In a recent overview paper \cite{barros20}, a perspective on future directions in applying \ac{MC} to viral infection research is given.
Finally, in \cite{koca21}, the \ac{sarscovtwo} transmission within the human respiratory tract is modeled using \ac{MC}.


In contrast to existing papers using \ac{MC} to study airborne virus transmission, we propose a novel model for airborne virus transmission via \acp{IA} that takes into account the randomness introduced by imperfectly fitted face masks.
Hence, in contrast to existing models, the model proposed in this paper allows for a comprehensive assessment of the infection risk imposed by imperfectly fitted face masks.
By numerical evaluation, we are able to show the impact of the physical face mask parameters on the infection probability of a \ac{HP} exposed to \acp{IA} emitted by an asymptomatic, regularly breathing \ac{IP}.
We further show the impact of the different breathing dynamics during exhalation and inhalation, respectively, on the stochastic filtering of \acp{IA} by the face mask.
Despite its simplicity, the proposed model generates insight into the statistics of the indoor transmission of \ac{sarscovtwo} from an infected, but asymptomatic person to a susceptible person both protected by imperfectly fitted face masks.

The remainder of this paper is organized as follows.
In Section~\ref{sec:physical_transmission_scenario}, we introduce the considered physical virus transmission scenario.
In Sections~\ref{sec:transmitter}, \ref{sec:channel}, and \ref{sec:receiver}, we present the transmitter, channel, and receiver models, respectively, required for framing airborne virus transmission as \ac{MC} system.
In Section~\ref{sec:numerical_results}, we present results from the numerical evaluation of the proposed model, before we summarize our main findings in Section~\ref{sec:conclusions}.

%% file: physical_transmission_scenario.tex
\vspace*{-0.5ex}
We consider a generic closed indoor environment with one \ac{HP} present.
Another person which is infected but asymptomatic, i.e., the \ac{IP}, is present in the room for $T_{\textrm{tot}}$ seconds.
After $T_{\textrm{tot}}$ seconds, the \ac{IP} leaves the room.
The \ac{HP} is present in the room while the \ac{IP} is present and stays in the room after the \ac{IP} has left.
Both the \ac{IP} and the \ac{HP} breath regularly while they are present in the room.
While breathing, the \ac{IP} exhales virus-carrying aerosols, i.e., the \acp{IA}\footnote{In general, the exhalation air contains aerosols of different sizes, some of which carry viruses and some of which do not. In this paper, we restrict our attention to small (average diameter $5$~\si{\micro\meter}), virus-carrying aerosols that remain dissolved in air for up to $2.6$ hours \cite{lelieveld20}.}.

Both the IP and the \ac{HP} wear face masks.
However, the masks do not act as perfect filters for the \acp{IA}, because
\begin{itemize}
    \item the face mask fabric does not block all \acp{IA} \cite{mittal20a} and
    \item the fit of the face masks to the faces of the \ac{IP} and the \ac{HP}, respectively, is not perfect, such that \acp{IA} possibly pass through gaps between the edge of the face mask and the skin during breathing \cite{lei12}.
\end{itemize}

\begin{figure*}
	\centering
	\includegraphics[width=\linewidth]{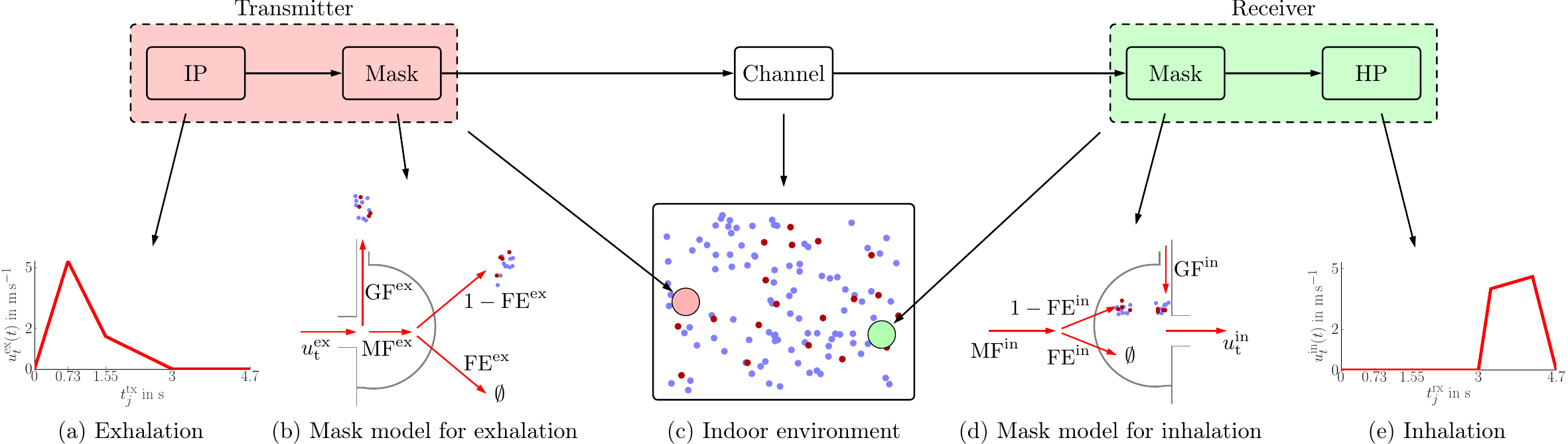}
	\caption{\small Block diagram of the considered transmission scenario: The \ac{HP} (green circle) is exposed to infectious aerosols (IAs) emitted by the \ac{IP} (red circle) in an indoor environment.
        The \ac{IP} breathes regularly and emits aerosols during exhalation (a), which can in general be non-infectious (blue dots) or infectious (red dots).
        While we consider only \acp{IA} in this paper, the non-infectious aerosols are included in this figure to show the relative abundance of non-infectious aerosols as compared to \acp{IA} in the exhaled air of the \ac{IP} \cite{cheng20}.
        Exhaled aerosols are emitted into the environment as part of the exhalation mask flow $\MFex$ through the face mask of the \ac{IP} or as part of the (unfiltered) exhalation gap flow $\GFex$ (b).
        The \ac{HP} breathes regularly and inhales (e) air.
        \acp{IA} captured by the inhalation air flow of the \ac{HP} are either part of the inhalation mask flow $\MFin$ or part of the (unfiltered) inhalation gap flow $\GFin$ (d).}
	\label{fig:process}
	\vspace*{-1.5ex}
\end{figure*}

When the \ac{IP} exhales, some of the viruses present in his respiratory tract dissolve in the exhalation air as \acp{IA} \cite{mittal20}.
Upon exhalation, some of these \acp{IA} deposit in the face mask of the \ac{IP}, some are released into the ambient air.
The exhaled \acp{IA} that did not deposit in the face mask of the \ac{IP} spread in the room due to the turbulent ambient air flow \cite{mittal20}.
Ultimately, each emitted \ac{IA} either deposits on a surface in the room or on the face mask of the \ac{IP} or the \ac{HP}, respectively, degrades with a rate depending on the room temperature and relative humidity, or is inhaled by the \ac{HP} or the \ac{IP} \cite{lelieveld20}.

In order to study the exposure of the \ac{HP} to \acp{IA} released by the \ac{IP}, we frame the airborne transmission scenario outlined above as \ac{MC} system, cf.~Fig.\ref{fig:process}.
To this end, we consider the \ac{IP} as molecular transmitter, the \ac{HP} as molecular receiver, and the room as channel.
In this system model, information is conveyed from the \ac{IP} to the \ac{HP} by the release of signaling molecules in the form of \acp{IA} into the channel.
The \ac{HP} repeatedly samples the \ac{IA} concentration in the environment by inhaling aerosols and accumulates the inhaled \acp{IA} in his respiratory tract.
If the cumulative number of inhaled \acp{IA} exceeds some threshold, the \ac{HP} gets infected, i.e., the virus has been transmitted from the \ac{IP} to the \ac{HP}.

Due to the impact of several random parameters such as the ambient air flow, the room temperature, and the air humidity on the propagation and degradation of \acp{IA} \cite{mittal20}, the virus transmission is random.
Furthermore, the face masks of the \ac{IP} and the \ac{HP}, respectively, render the release and the inhalation of the \acp{IA} random, as some of the \acp{IA} are filtered by the masks.
In the following sections, we present models for the transmitter, the channel, and the receiver, respectively, that encompass this randomness and finally allow us to evaluate the statistics of the number of \acp{IA} inhaled by the \ac{HP}.

%% file: transmitter.tex
\vspace*{-0.5ex}
In the considered airborne virus transmission system, cf.~Section~\ref{sec:physical_transmission_scenario}, virions deposited in the respiratory tract of the \ac{IP} dissolve in the exhalation air flow and are transported as \acp{IA} through the \ac{IP}'s mouth/nose.
After leaving the mouth/nose of the IP, the IAs are either filtered by the face mask of the IP (and not further considered in our model) or emitted into the environment (physical channel).
Hence, the molecular transmitter 
in our model involves three sequential stages:
\begin{itemize}
    \item Generation of solute \acp{IA} within the respiratory tract during exhalation
    \item Exhalation of \acp{IA} through the mouth/nose into the volume enclosed by the face mask and the face of the \ac{IP}
    \item Deposition of \acp{IA} in the face mask or release into the environment
\end{itemize}

\subsection{Number of Exhaled Infectious Aerosols}
\label{sec:transmitter:number_of_EIAs}
\vspace*{-0.5ex}
We assume that the \ac{IP} stays in the room for $T_{\textrm{tot}}$ seconds.
For simplicity, we also assume that $T_{\textrm{tot}}$ is an integer multiple of $T$, i.e., $T_{\textrm{tot}}/T = I \in \mathbb{N}$, where $\mathbb{N}$ denotes the set of positive integers and $I$ denotes the number of breaths that the \ac{IP} takes in the room.
For each interval $i \in \lbrace 1,\ldots,I \rbrace$, we denote the number of exhaled IAs in $i$ by $N_i$.
According to \cite{cheng20}, $N_i$ follows a Poisson distribution with mean $\lambda$, i.e., $N_i \sim \textrm{Pois}(\lambda)$, and, according to empirical observations, $\lambda$ ranges from $0.0000000049$ virus copies per \si{\cubic\centi\meter} for low emitters to $0.637$ virus copies per \si{\cubic\centi\meter} for high emitters infected with \ac{COVID-19} \cite{riediker20}.
Finally, we denote the set of \acp{IA} exhaled during $T_{\textrm{tot}}$ as $\mathcal{M} = \lbrace M_1,\ldots,M_J \rbrace$, where $J = \sum_{i = 1}^{I} N_i$.
Since the $N_i$ are independent Poisson \acp{RV} with identical mean $\lambda$, $J$ is a Poisson \ac{RV} with mean $\lambda I$, i.e., $J \sim \textrm{Pois}(\lambda I)$.

\subsection{Exhalation Model}
\vspace*{-0.5ex}
We assume that the regular breathing of the \ac{IP} follows the experimentally obtained breathing pattern reported in \cite{lei12,russo11}.
In this model, one breath consists of one exhalation followed by one inhalation and regular breathing is a sequence of identical breaths.
This means that each breath in such a sequence of breaths is identical with respect to its total duration and the relative duration of exhalation and inhalation.
Let $T$ denote the length of one breath in seconds.
Let further denote $t^{\textrm{tx}}_j$, $j \in \lbrace 1, \ldots, J \rbrace$, the time variable ranging over the interval $i \in \lbrace 1,\ldots,I \rbrace$ in which \ac{IA} $j$ is exhaled {\em relative to the beginning} of this interval, i.e., $t^{\textrm{tx}}_j \in [0,T]$.
For regular breathing, $T=\SI{4.7}{\second}$ according to \cite{lei12,russo11}.
Hence, during breath $i$ of the \ac{IP}, the absolute velocity of the exhalation air flow, $u^{\textrm{ex}}_t(t^{\textrm{tx}}_j)$ in $\si{\meter\per\second}$, is then given by \cite{lei12,russo11}
\vspace{-0.5ex}
\begin{align}
    &u^{\textrm{ex}}_t(\ttx_j) =\nonumber\\
    &\begin{cases}
        \frac{5.3}{0.73}\frac{\si{\meter}}{\si{\second\squared}} \ttx_j, \quad & 0\,\si{\second} \leq \ttx_j < 0.73\,\si{\second}, \\
        \frac{(1.6-5.3)}{(1.55-0.73)}\frac{\si{\meter}}{\si{\second\squared}} (\ttx_j-0.73\,\si{\second}) + 5.3 \frac{\si{\meter}}{\si{\second}}, \quad & 0.73\,\si{\second} \leq \ttx_j < 1.55\,\si{\second}, \\
        \frac{-1.6}{(3-1.55)}\frac{\si{\meter}}{\si{\second\squared}} (\ttx_j-1.55\,\si{\second}) + 1.6 \frac{\si{\meter}}{\si{\second}}, \quad & 1.55\,\si{\second} \leq \ttx_j < 3\,\si{\second}, \\
        0 \frac{\si{\meter}}{\si{\second}}       , \quad & 3\,\si{\second} \leq \ttx_j < 4.7\,\si{\second}.
    \end{cases}\label{eq:exhalation_fct}
\end{align}
According to \eqref{eq:exhalation_fct}, $\uex_t(\ttx_j)$ is a piecewise linear function defined for $\ttx_j \in [0,T]$.
Fig.~\ref{fig:process}(a) shows $u^{\textrm{ex}}_t(\ttx_j)$ as defined in \eqref{eq:exhalation_fct}.

\subsection{Exhalation Time}
\vspace*{-0.5ex}
After an \ac{IA} $M_j$, $j \in \lbrace 1,\ldots, J \rbrace$, has been exhaled in interval $i$, it is either captured by the face mask of the \ac{IP} or released into the room.
We recall from Section~\ref{sec:physical_transmission_scenario} that the face mask is imperfectly fitted to the face of the \ac{IP} and, hence, some of the exhaled aerosols leak through gaps between the face mask and the skin without passing through the mask fabric.
This leakage flow, in turn, depends on the velocity of the exhalation air flow \cite{peric20} at the time at which $M_j$ is exhaled.
Now, assuming that \acp{IA} are uniformly distributed over the total volume of air exhaled in breath $i$, the probability that $M_j$ is exhaled in any sub-interval of this breath is proportional to the exhalation air flow rate, i.e.,
\vspace{-0.5ex}
\begin{align}
    \textrm{Pr}\lbrace T^{\textrm{ex}}_j \leq \ttx_j \rbrace = \frac{\int_{0}^{\ttx_j} \Fex_t(\tau) \dtau}{\int_{0}^{T} \Fex_t(\tau) \dtau} = \frac{\int_{0}^{\ttx_j} \uex_t(\tau) \dtau}{\int_{0}^{T} \uex_t(\tau) \dtau}, 0 \leq \ttx_j \leq T,\label{eq:exhalation_time}
\end{align}
where $T^{\textrm{ex}}_j$ denotes the random time of exhalation of $M_j$ relative to the beginning of breath $i$ and $\Fex_t(\ttx_j)$ denotes the total flow rate of exhaled air in $\si{\cubic\meter\per\second}$ at $t^{\textrm{tx}}_j$, i.e., $\Fex_t(t^{\textrm{tx}}_j) = \uex_t(t^{\textrm{tx}}_j) \Stx_t$, where $\Stx_t$ denotes the surface area of the mouth/nose of the IP in $\si{\square\meter}$.
According to \eqref{eq:exhalation_time}, $\textrm{Pr}\lbrace T_j \leq t^{\textrm{tx}}_j\rbrace$ can 
be obtained by integrating \eqref{eq:exhalation_fct} (which is straightforward).

\subsection{Escape Probability}\label{sec:transmitter:escape_propability}
\vspace*{-0.5ex}
To account for the imperfect fit of the face mask of the \ac{IP}, we distinguish the air flows through the face mask and through the gaps at the edge of the mask during exhalation and call them {\em exhalation mask flow (\MFex)} and {\em exhalation gap flow (\GFex)}, respectively.
To characterize \MFex and \GFex, we adopt the analytical fluid dynamics model for the breathing of a person through a face mask from \cite{peric20}.
In this simplified model, the face area around the mouth/nose of the person is modeled as a planar surface, the face mask is modeled as a half-sphere, and the gaps are modeled as one cuboid extending from the surface of the mask towards the environment, see~Fig.~\ref{fig:process}(b) for a schematic representation of the model and \cite[Fig.~1]{peric20} for details.
The model from \cite{peric20} quantifies \MFex and \GFex as functions of various parameters of the face mask and the exhaled air flow rate.
Furthermore, it has been validated by three-dimensional fluid dynamics simulation in \cite{peric20}.

Let $\Fex_g(t)$ denote the air flow rate through the gaps at the edge of the face mask of the IP in $\si{\cubic\meter\per\second}$ at time $t$, $0 \leq t \leq T$.
According to \cite{peric20} and assuming spatially uniform air pressure within the volume enclosed by the face mask and the skin of the IP at any time $t$, $\Fex_g(t)$ is given by \cite{peric20}
\begin{align}
    \Fex_g(t) = \uex_g(t) \Htx_g  \Btx_g, \label{eq:gap_flow}
\end{align}
where $\Htx_g$ and $\Btx_g$ denote the height and the width, respectively, of the gap in $\si{\meter}$ and $\uex_g(t)$ is given by \cite{peric20}
\vspace*{-1ex} 
\begin{align}
    \uex_g(&t) = \frac{1}{\zeta \rho}\left[-\left(\frac{12 \mu \Ltx_g}{\left( \Htx_g \right)^2} + \frac{\Ctx_m \rho \Htx_g \Btx_g}{\Stx_m}\right) +\right.\nonumber\\
    {}\!\!&+\left.\!\!\sqrt{\left(\frac{12 \mu \Ltx_g}{\left( \Htx_g \right)^2} \!+\! \frac{\Ctx_m \rho \Htx_g \Btx_g}{\Stx_m}\right)^2 \!\!\!\!+\! 2 \frac{\zeta \rho^2 \Ctx_m \Fex_t(t)}{\Stx_m}}\right],\label{eq:gap_flow_velocity}
\end{align}
where the constants $\Ltx_g$, $\Ctx_m$, $\Stx_m$, $\zeta$, $\mu$, and $\rho$ denote the length of the gap in $\si{\meter}$, the viscous porous resistance of the face mask of the \ac{IP} in $\si{\meter\per\second}$, the surface area of the face mask of the \ac{IP} in $\si{\square\meter}$, the pressure loss coefficient at the gap in $\si{\kilogram\per\meter}$, the dynamic viscosity of air in $\si{\pascal\second}$, and the density of air in $\si{\kilogram\per\cubic\meter}$, respectively.
Hence, $\Ltx_g$ depends on the fit of the mask to the face of the \ac{IP}, $\Ctx_m$ and $\Stx_m$ are properties of the face mask of the \ac{IP}, and $\zeta$, $\mu$, and $\rho$ are physical constants.

From \eqref{eq:gap_flow} and $\Fex_t$, we determine the probability that $M_j$, exhaled at time $T^{\textrm{ex}}_j$, escapes through the gap as
\vspace*{-1ex} 
\begin{align}
    \textrm{Pr}\lbrace M_j \in \GFex \rbrace = \frac{\Fex_g(T^{\textrm{ex}}_j)}{\Fex_t(T^{\textrm{ex}}_j)} = 1 - \textrm{Pr}\lbrace M_j \in \MFex \rbrace.\label{eq:prob_gap_ex}
\end{align}
If \ac{IA} $M_j$ instead is carried by the mask flow $\MFex$, it deposits in the mask with a probability corresponding to the outward filtration efficiency of the mask, $\FEex$, and passes the mask unfiltered with probability $1-\FEex$ \cite{mittal20a}.
Hence, the emission of $M_j$ into the environment can be modeled as a Bernoulli \ac{RV} $\Stx_j \in \lbrace 0, 1 \rbrace$, with probabilities
\vspace*{-0.5ex} 
\begin{align}
    \Pr\lbrace \Stx_j \!&= \!1 \rbrace = 1 \!-\! \Pr\lbrace \Stx_j = 0 \rbrace\nonumber\\
    &= \Pr\lbrace M_j \!\in \GFex \rbrace \!+\! \textrm{Pr}\lbrace M_j \!\in \MFex\rbrace (1-\FEex),
\end{align}
where $\Stx_j = 1$ and $\Stx_j = 0$ denote the events that $M_j$ is emitted into the environment and $M_j$ is filtered by the mask, respectively.

Finally, the total number of \acp{IA} emitted into the environment by the \ac{IP} is given by
\vspace*{-3ex} 
\begin{align}
    \EIA = \sum_{j=1}^{J} \Stx_j,\label{eq:no_emitted_ia}
\end{align}
$\EIA \in \lbrace 0,\ldots, J \rbrace$, and we denote the emitted \acp{IA} by $E_k$, $k \in$ ${}$ $\lbrace k_1, \ldots, k_{\EIA} \rbrace \subseteq \lbrace 1, \ldots, J \rbrace$.
According to \eqref{eq:no_emitted_ia}, the discrete \ac{RV} $\EIA$ is the sum of $J$ independent Bernoulli \acp{RV} with different success probabilities depending on the random exhalation times of \acp{IA} $M_j$.
If the ratio of the flow rates of $\GFex$ and $\MFex$ was constant with respect to time, i.e., $\Fex_g(t_1)/\Fex_t(t_1) = \Fex_g(t_2)/\Fex_t(t_2) = q$, for all $t_1,t_2 \in [0,T^{\textrm{ex}}_j]$, $\EIA$ would be a binomial distributed \ac{RV} with success probability $q$.
However, due to the nonlinear relationship between $\Fex_g(t)$ and $\Fex_t(t)$ in \eqref{eq:gap_flow}, \eqref{eq:gap_flow_velocity}, this is not the case.

%% file: channel.tex
\vspace*{-0.5ex}
According to Section~\ref{sec:physical_transmission_scenario}, after an IA $E_k$ is emitted into the environment by the IP, it is subject to the turbulent flow of the ambient air before it either deposits on a surface, for example on a wall or on furniture, or is inhaled by either the \ac{IP} or the \ac{HP}.
According to \cite{mittal20a}, the probability that $E_k$ arrives at the \ac{HP} in a viable, i.e., infectious, state can be written as the product of two probabilities, $p_a p_v$, where $p_a \in [0,1]$ denotes the probability that $E_k$ reaches the \ac{HP} and $p_v \in [0,1]$ denotes the probability that the virion carried by the aerosol is still infectious when reaching the \ac{HP}.
Depending on the room geometry, furniture, ventilation, and the positions of the \ac{IP} and the \ac{HP} in the room, $p_a$ can be estimated using tools from fluid mechanics \cite{bhagat20}.
$p_v$ depends on environmental conditions such as the relative humidity and temperature, on the travel time from the \ac{IP} to the \ac{HP}, and on the half-time of \ac{sarscovtwo} virions on aerosols \cite{mittal20a}.
Now we define the Bernoulli \ac{RV} $\Sch_k \in \lbrace 0, 1 \rbrace$ to indicate whether $E_k$ reaches the \ac{HP} in a viable state and, recalling that we have assumed that $E_k$ did not deposit in the face mask of the \ac{IP} during exhalation, we obtain
\begin{align}
    \textrm{Pr}\lbrace \Sch_k =\! 1 | \Stx_k =\! 1 \rbrace = p_a p_v = 1 \!-\! \textrm{Pr}\lbrace \Sch_k = 0 | \Stx_k = 1 \rbrace,
\end{align}
where $\Sch_k = 1$ and $\Sch_k = 0$ denote the events that $E_k$ is captured and not captured by the inhalation air flow of the \ac{HP} in a viable state, respectively, and $p_a p_v$ is identical for all \acp{IA}.
Hence, without making any further assumptions on the room geometry, ventilation, and other properties of the physical environment, the propagation of $E_k$ from the \ac{IP} to the \ac{HP} is modeled as Bernoulli \ac{RV}\footnote{Modeling molecule propagation as Bernoulli \ac{RV} has been suggested in the context of airborne virus transmission \cite{mittal20a} and has also been successfully applied in the context of \ac{MC} \cite{chahibi14}.}.
Due to the block-based approach used in this paper (see Fig.~\ref{fig:process}), 
more detailed propagation models can be readily adopted.
While this is certainly an interesting extension to the model proposed in this paper, due to the space constraints, this is left for future work.

%% file: receiver.tex
\vspace*{-0.5ex}
In this section, we present a detailed model for the inhalation of \acp{IA} by the \ac{HP} given that the \ac{HP} wears an imperfectly fitted face mask that filters part of the aerosols from the inhaled ambient air.
The presented model parallels in part the model presented in Section~\ref{sec:transmitter}.

\subsection{Inhalation Model}
\vspace*{-0.5ex}
Analogously to Section~\ref{sec:transmitter}, let $\trx_j \in [0,T]$ denote the time variable ranging over the interval in which \ac{IA} $j$ is captured by the inhalation air flow of the \ac{HP} (this does not imply that $j$ is actually {\em inhaled} by the \ac{HP}, as there is still a chance that it deposits in the \ac{HP}'s face mask) {\em relative to the beginning} of this interval, i.e., $\trx_j \in [0,T]$.
The inhalation of the \ac{HP} in this interval is described by the time-dependent inhalation air flow velocity $\uin_t(\trx_j)$ in $\si{\meter\per\second}$ given by \cite{lei12,russo11}
\begin{align}
    &\uin_t(\trx_j) = \nonumber\\
    &\begin{cases}
        0 \frac{\si{\meter}}{\si{\second}}       , \quad & 0\,\si{\second} \leq \trx_j < 3\,\si{\second},\\
        \frac{4}{(3.27-3)}\frac{\si{\meter}}{\si{\second\squared}} (\trx_j-3\,\si{\second}), \quad & 3\,\si{\second} \leq \trx_j < 3.27\,\si{\second}, \\
        \frac{(4.6-4)}{(4.18-3.27)}\frac{\si{\meter}}{\si{\second\squared}} (\trx_j-3.27\,\si{\second}) + 4 \frac{\si{\meter}}{\si{\second}}, \quad & 3.27\,\si{\second} \leq \trx_j < 4.18\,\si{\second}, \\
        \frac{-4.6}{(4.7-4.18)}\frac{\si{\meter}}{\si{\second\squared}} (\trx_j-4.18\,\si{\second}) + 4.6 \frac{\si{\meter}}{\si{\second}}, \quad & 4.18\,\si{\second} \leq \trx_j < 4.7\,\si{\second}.
    \end{cases}\label{eq:inhalation_fct}
\end{align}
Eq.~\eqref{eq:inhalation_fct} is illustrated in Fig.~\ref{fig:process}(e)\footnote{Note that we do not assume that the \ac{IP} and the \ac{HP} start breathing at the same time.}.

\subsection{Inhalation Time}
\vspace*{-0.5ex}
We assume that the IA $E_k$ reaches the \ac{HP} in a viable state according to the definition in Section~\ref{sec:channel}, i.e., $\Sch_k = 1$.
Then, the probability that $E_k$ is carried by the inhalation air flow towards the respiratory tract of the \ac{HP} at any random time $\Tin_k \in [0,T]$ is proportional to the inhalation air flow, i.e.,
\begin{align}
    \textrm{Pr}\lbrace \Tin_k \leq \trx_j\rbrace = \frac{\int_{0}^{\trx_j} \Fin_t(\tau) \dtau}{\int_{0}^{T} \Fin_t(\tau) \dtau} = \frac{\int_{0}^{\trx_j} \uin_t(\tau) \dtau}{\int_{0}^{T} \uin_t(\tau) \dtau}, 0 \leq \trx_j \leq T,\label{eq:inhalation_time}
\end{align}
where the total inhalation air flow rate at time $\trx_j$, $\Fin_t(\trx_j)$ in \si{\meter\cubed\per\second}, is defined as
\vspace*{-1.5ex}
\begin{align}
    \Fin_t(\trx_j) = \uin_t(\trx_j) \Srx_t,
\end{align}
where $\Srx_t$ denotes the surface area of the mouth/nose of \ac{HP} in $\si{\square\meter}$.

\subsection{Inhalation Mask Flow and Gap Flow}
\vspace*{-0.5ex}
Since the \ac{HP}, similar to the \ac{IP}, wears an imperfectly fitted face mask, the air flow resulting from the inhalation of the \ac{HP} is split into a mask flow component ($\MFin$) and a gap flow component ($\GFin$).
To model $\MFin$ and $\GFin$, as in Section~\ref{sec:transmitter}, we adopt the fluid dynamics model from \cite{peric20}.
When an \ac{IA} $E_k$ reaches the \ac{HP}, it follows the $\GFin$ with a probability proportional to the flow through the gap relative to the total inhalation flow at time $\Tin_k$, i.e.,
\vspace*{-1ex}
\begin{align}
    \textrm{Pr}\lbrace E_k \in \GFin \rbrace = \frac{\Fin_g(\Tin_k)}{\Fin_t(\Tin_k)} = 1 - \textrm{Pr}\lbrace E_k \in \MFin \rbrace,\label{eq:prop_gf_in}
\end{align}
where the air flow through the gap at time $t \in [0,T]$, $\Fin_g(t)$ in \si{\meter\cubed\per\second}, is defined as
\vspace*{-0.5ex}
\begin{align}
    \Fin_g(t) = \uin_g(t) \Hrx_g  \Brx_g,
\end{align}
where $\Hrx_g$ and $\Brx_g$ denote the height and the width of the gap between the face mask and the skin of the HP in $\si{\meter}$, respectively.
The flow velocity of the gap flow, $\uin_g(t)$ in $\si{\meter\per\second}$ is given by \cite{peric20}
\vspace*{-1ex}
\begin{align}
    \uin_g(&t) = \frac{1}{\zeta \rho}\left[-\left(\frac{12 \mu \Lrx_g}{\left( \Hrx_g \right)^2} + \frac{\Crx_m \rho \Hrx_g \Brx_g}{\Srx_m}\right) +\right.\nonumber\\
    {}\!\!\!&+\left.\!\sqrt{\left(\frac{12 \mu \Lrx_g}{\left( \Hrx_g \right)^2} \!+\! \frac{\Crx_m \rho \Hrx_g \Brx_g}{\Srx_m}\right)^2 \!\!\!\!-\! 2 \frac{\zeta \rho^2 \Crx_m \Fin_t(t)}{\Srx_m}}\right]\!\!.\label{eq:inhalation_gap_flow_velocity}
\end{align}
Analogous to Section~\ref{sec:transmitter:escape_propability}, the constants $\Lrx_g$, $\Crx_m$, and $\Srx_m$ denote the length of the gap in $\si{\meter}$, the viscous porous resistance of the \ac{HP}'s face mask in $\si{\meter\per\second}$, and the surface area of the \ac{HP}'s face mask in $\si{\square\meter}$, respectively, and $\zeta$, $\mu$, and $\rho$ are defined in Section~\ref{sec:transmitter:escape_propability}.
Eq.~\eqref{eq:prop_gf_in} follows from the observation that every inhaled aerosol is either part of $\GFin$ or
$\MFin$.

\subsection{Inhalation Probability}
\vspace*{-0.5ex}
With the observations from the previous section, we are able to determine the probability for any \ac{IA} $E_k$ reaching the \ac{HP} at time $\Tin_k$ to actually pass the face mask and reach the respiratory tract of the \ac{HP}.
Similar to Section~\ref{sec:transmitter:escape_propability}, we model the inhalation of $E_k$ as a binary RV $\Srx_k \in \lbrace 0, 1 \rbrace$, where $\Srx_k = 1$ and $\Srx_k = 0$ denote the events that $E_k$ is inhaled by the \ac{HP} and $E_k$ is filtered by the face mask of the \ac{HP}, respectively.
Now, $\Srx_k = 1$ if and only if $E_k$ is part of $\GFin$ {\em or} $E_k$ is part of $\MFin$ and not filtered by the face mask.
Then, recalling that we have assumed that $E_k$ did not deposit in the face mask of \ac{IP} and reached \ac{HP}, we obtain
\begin{align}
    \Pr \lbrace \Srx_k &= 1 | \Stx_k = 1, \Sch_k = 1 \rbrace\nonumber\\
    &= 1 - \Pr\lbrace \Srx_k = 0 | \Stx_k = 1, \Sch_k = 1 \rbrace\nonumber\\
    &= \Pr \lbrace E_k \!\in \GFin \rbrace \!+\! \Pr\lbrace E_k \!\in \MFin \rbrace (1\!-\!\FEin),
\end{align}
and $\FEin$ is the inward filtration efficiency of the face mask of the \ac{HP}.

\subsection{Detection}
\vspace*{-0.5ex}
Let the binary RV $S_j$ denote the event that IA $j$ is transmitted from \ac{IP} to \ac{HP} via airborne transmission, i.e., $S_j = 1$ if \ac{IA} $j$ is transmitted and $S_j = 0$ otherwise.
According to the results presented in the previous sections, we can write $S_j$ as
\vspace*{-0.5ex}
\begin{align}
    S_j = \Stx_j \Sch_j \Srx_j,\label{eq:end-to-end_transmission}
\end{align}
where
\vspace*{-1ex}
\begin{align}
    \Pr\lbrace S_j = 1 \rbrace &= \Pr\lbrace\Stx_j = 1, \Sch_j = 1, \Srx_j = 1 \rbrace\nonumber\\
    &= \Pr\lbrace\Stx_j = 1\rbrace \Pr\lbrace \Sch_j = 1 | \Stx_j = 1\rbrace\times\nonumber\\
    &\times\Pr\lbrace\Srx_j = 1 | \Stx_j = 1, \Sch_j = 1 \rbrace\nonumber\\
    &= 1 - \Pr\lbrace S_j = 0 \rbrace.
\end{align}
The random number of inhaled \acp{IA} by \ac{HP}, $\IIA$, is given by
\vspace*{-1ex}
\begin{align}
    \IIA = \sum_{j = 1}^{J} S_j.\label{eq:def_iia}
\end{align}
According to \eqref{eq:def_iia}, $\IIA$ is a discrete random variable with support $\mathbb{N}_0$, where $\mathbb{N}_0$ denotes the set of non-negative integers.
Note that the $S_j$ are independent Bernoulli RVs, but they are not identically distributed due to their dependence on the random exhalation and inhalation times of each \ac{IA}, respectively.
Hence, $\IIA$ does in general not follow a Binomial distribution.

According to commonly used models for virus transmission, \ac{HP} gets infected if $\IIA$ reaches the infectious dose $\theta$, $\theta \in \mathbb{N}_0$, that depends on the physical state and the state of the immune system of the \ac{HP} \cite{mittal20a}.
Hence, the virus is transmitted from \ac{IP} to \ac{HP} with probability $P(\theta)$, where
\vspace*{-2.5ex}
\begin{align}
    P(\theta) = \Pr\lbrace \IIA \geq \theta \rbrace = 1 - \sum_{n = 0}^{\theta-1} \Pr\lbrace \IIA = n \rbrace.\label{eq:infection_probability}
\end{align}
We refer to $P(\theta)$ as {\em infection probability}.

%% file: numerical_results.tex
\vspace*{-0.5ex}
In this section, we present numerical results for the \ac{MC} based airborne virus transmission model presented in Sections~\ref{sec:transmitter}, \ref{sec:channel}, and \ref{sec:receiver}.
To this end, 
we first describe our simulation framework and the parameters used.
Then, we study the impact of some of these parameters on the end-to-end system.

\subsection{Simulation and Choice of Parameters}
\vspace*{-0.5ex}
To simulate the transmission scenario described in Sections~\ref{sec:transmitter}, \ref{sec:channel}, and \ref{sec:receiver}, we first computed the random number of exhaled IAs, $J$, according to Section~\ref{sec:transmitter:number_of_EIAs}.
Next, we simulated the transmission of each of the $J$ IAs from IP to HP as a sequence of random experiments according to \eqref{eq:end-to-end_transmission}.
To this end, we first computed the random exhalation and inhalation times for each IA using inverse transform sampling \cite{devroye86} after integrating and inverting \eqref{eq:exhalation_time} and \eqref{eq:inhalation_time}.
Then, we used \eqref{eq:end-to-end_transmission} to determine for each of the $J$ exhaled IAs whether it was inhaled by the HP or not and finally summed up the inhaled IAs to obtain the random number of inhaled infectious aerosols, IIA, according to \eqref{eq:def_iia}.
We repeated this procedure $10,000$ times to obtain the empirical distribution of the RV IAA as presented in the following sections.
%
Table~\ref{tab:sim_params} lists the default simulation parameters.

\begin{table}
    \vspace*{0.07in}
    \centering
    \caption{Default Simulation Parameters.}
    \vspace*{-0.02in}
    \footnotesize
    \begin{tabular}{| p{.12\linewidth} | r | p{.39\linewidth} | c |}
        \hline Parameter & Default Value & Description & Ref.\\ \hline
        $\Htx_g$, $\Hrx_g$ & $\SI{1e-3}{\meter}$ & Face mask gap height & \cite{peric20}\\ \hline
        $\Btx_g$, $\Brx_g$ & $\SI{2.5e-2}{\meter}$ & Face mask gap width & \cite{peric20}\\ \hline
        $\Ltx_g$, $\Lrx_g$ & $\SI{7.1e-3}{\meter}$ & Face mask gap length & \cite{peric20}\\ \hline
        $\Ctx_m$, $\Crx_m$ & $\SI{2e3}{\meter\per\second}$ & Face mask viscous porous resistance & \cite{peric20}\\ \hline
        $\Stx_m$, $\Srx_m$ & $\SI{1.58e-2}{\square\meter}$ & Face mask surface area & \cite{peric20}\\ \hline
        $\Stx_t$, $\Srx_t$ & $\SI{1e-4}{\square\meter}$ & Respiratory tract surface area & \cite{peric20}\\ \hline
        $\FEex$, $\FEin$ & $0.95$ & Face mask filtration efficiency & \\ \hline
        $\zeta$ & $\SI{1.5}{\kilogram\per\meter}$ & Pressure loss coefficient & \cite{peric20}\\ \hline
        $\mu$ & $\SI{1.8e-5}{\pascal\second}$ & Dynamic viscosity of air & \cite{peric20}\\ \hline
        $\rho$ & $\SI{1.2}{\kilogram\per\cubic\meter}$ & Density of air & \cite{peric20}\\ \hline
        $\lambda$ & $0.5$ & Average number of exhaled IAs per breath & \cite{riediker20}\\ \hline
        $I$ & $360$ & Number of breaths of IP & \\ \hline
        $p_a p_v$ & $0.1$ & 
		Channel transmission probability         
        & \\ \hline
    \end{tabular}\vspace*{-5mm}
    \label{tab:sim_params}
\end{table}

\subsection{Leakage at IP's Face Mask}
\vspace*{-0.5ex}
First, we examined how the fit of the face mask of the \ac{IP} impacts the probability that an exhaled \ac{IA} passes through the gap.
To this end, we drew $5000$ random exhalation times according to \eqref{eq:exhalation_time} and computed the respective random probabilities that particles exhaled at these time instants are part of the gap flow $\GFex$, cf.~\eqref{eq:prob_gap_ex}.

\begin{figure}[!t]
    \centering
    \includegraphics[width=.42\textwidth]{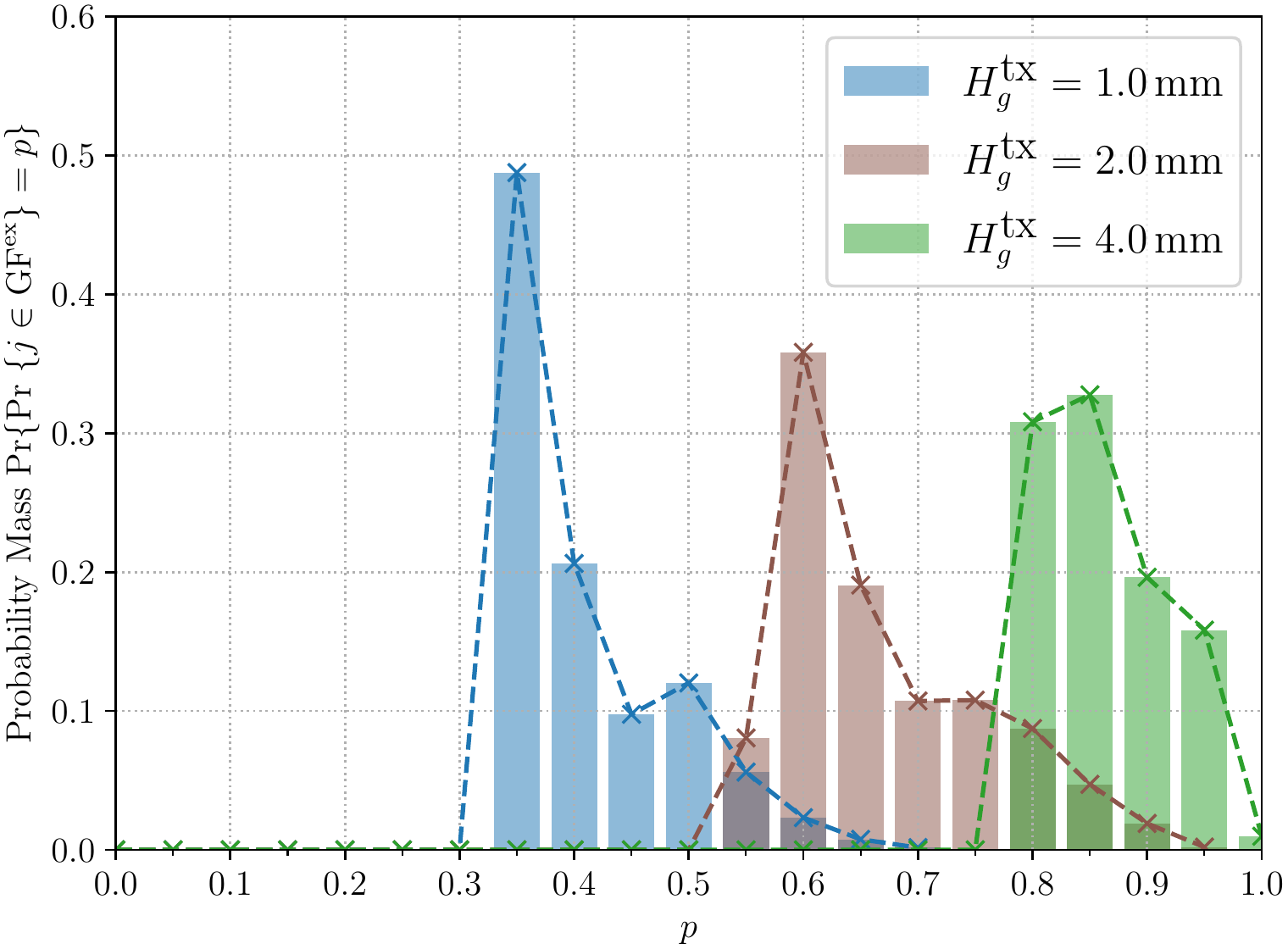}
    \caption{\small Empirical distribution of the probability that an \ac{IA} passes through the gap at IP's face mask during exhalation for different gap heights $\Htx_g$.}\vspace*{-4 mm}
    \label{fig:exhalation_gap_flow_probability_pmf}
\end{figure}

\begin{figure}[!t]
    \centering
    \includegraphics[width=.42\textwidth]{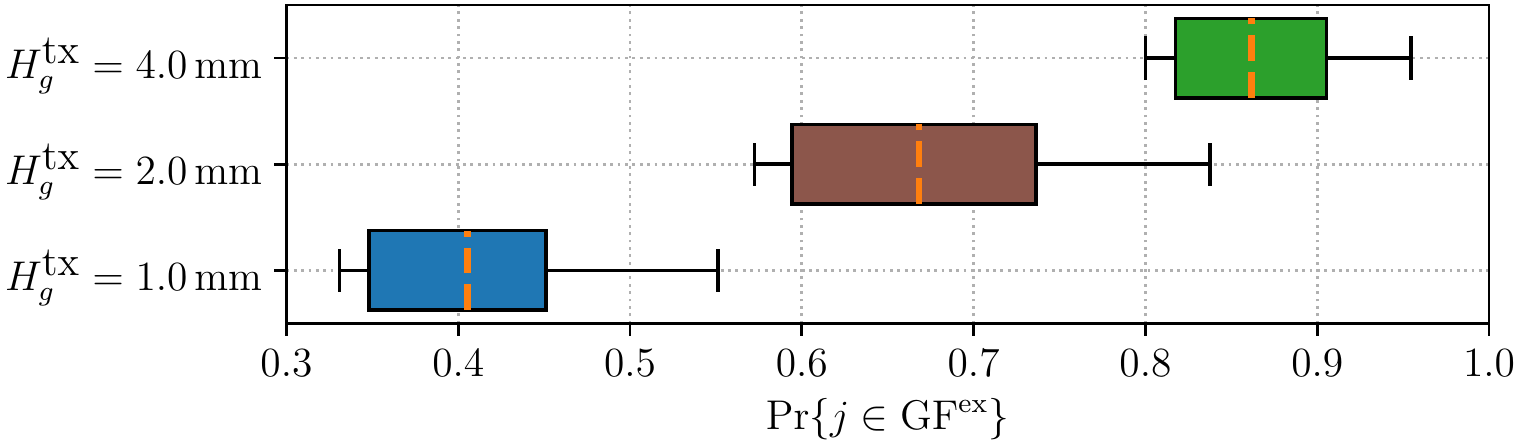}
    \caption{\small Empirical distribution of the probability that an \ac{IA} passes through the gap at IP's face mask during exhalation for different gap heights $\Htx_g$. The orange dashed lines indicate the empirical mean value. The left and right ends of the boxes indicate the third and first quartile, respectively. The left and right whiskers end at the 95\% and the 5\% quantile, respectively.}\vspace*{-4 mm}
    \label{fig:exhalation_gap_flow_probability_boxplot}
\end{figure}

In Fig.~\ref{fig:exhalation_gap_flow_probability_pmf}, we show the empirical frequencies of these probabilities after binning them into bins of width $0.05$ for different gap heights $\Htx_g$.
We observe from Fig.~\ref{fig:exhalation_gap_flow_probability_pmf} that the mean probability for an exhaled particle to escape through the gap increases as we increase $\Htx_g$.
Furthermore, we observe that the random escape probabilities spread considerably.
For example, for the default parameter setting from Table~\ref{tab:sim_params} shown in blue in Fig.~\ref{fig:exhalation_gap_flow_probability_pmf}, the escape probabilities are spread over the interval $[0.35;0.65]$.
In Fig.~\ref{fig:exhalation_gap_flow_probability_boxplot}, the empirical frequencies of the escape probabilities are shown as box plots.
In this representation, we can directly compare the distributions of the escape probabilities for different values of $\Htx_g$.
In particular, we observe from Fig.~\ref{fig:exhalation_gap_flow_probability_boxplot} that the spread of the empirical escape probabilities is relatively large for $\Htx_g = \SI{1}{\milli\meter}$ and $\Htx_g = \SI{2}{\milli\meter}$ and decreases only for $\Htx_g = \SI{4}{\milli\meter}$ as in this case almost all particles escape through the gap.

\subsection{Leakage at HP's Face Mask}
\vspace*{-0.5ex}
Next, we examine the randomness contributed by the splitting of the inhalation air flow into mask flow and gap flow at the \ac{HP}.
To this end, similar to the previous section, we drew $5000$ random inhalation times according to \eqref{eq:inhalation_time} and computed the respective probabilities that an \ac{IA} passes through the gap at the \ac{HP}'s face mask.
Then, we studied the empirical distribution of these probabilities for different viscous porous resistances $\Crx_m$ of the \ac{HP}'s face mask.

\begin{figure}[!t]
    \centering
    \includegraphics[width=.42\textwidth]{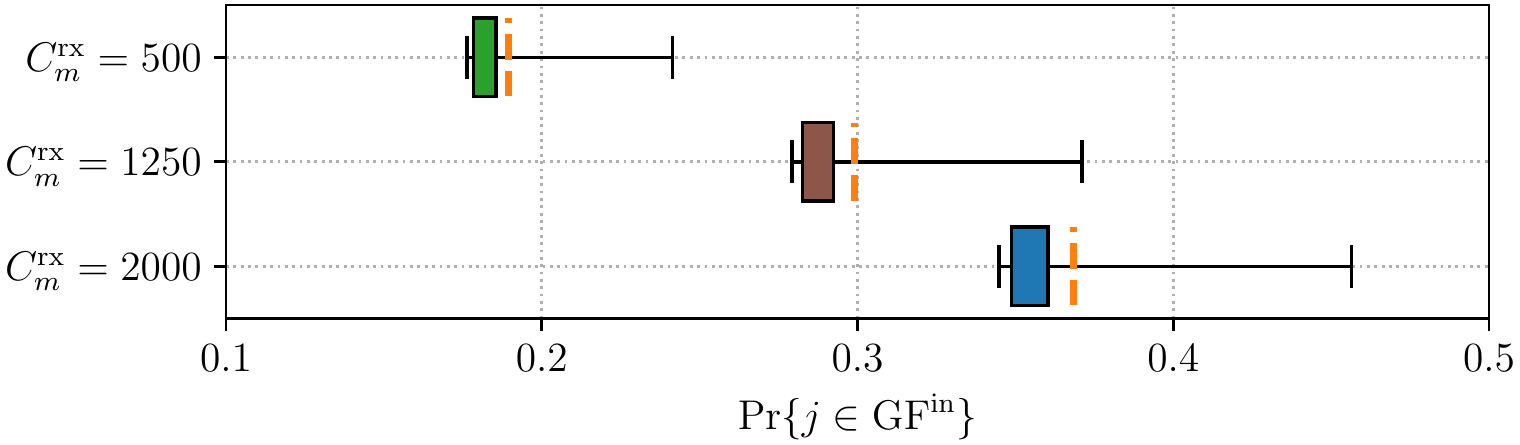}
    \caption{\small Empirical distribution of the probability that an \ac{IA} passes through the gap at HP's face mask during inhalation for different viscous porous resistances $\Crx_m$. The orange dashed lines indicate the empirical mean value. The left and right ends of the boxes indicate the third and first quartile, respectively. The left and right whiskers end at the 95\% and the 5\% quantile, respectively.}\vspace*{-4 mm}
    \label{fig:inhalation_gap_flow_probability_boxplot}
\end{figure}

From Fig.~\ref{fig:inhalation_gap_flow_probability_boxplot}, we observe that the spread of these probabilities is rather small as compared to the exhalation case, cf.~Fig.~\ref{fig:exhalation_gap_flow_probability_boxplot}.
Furthermore, we observe from Fig.~\ref{fig:inhalation_gap_flow_probability_boxplot} that the distribution of the probabilities remains concentrated even as $\Crx_m$ decreases, i.e., in very different regimes.
Indeed, this observation is plausible given that the inhalation flow as determined by \eqref{eq:inhalation_fct} exhibits a smaller dynamic range than the exhalation flow, cf.~Fig.~\ref{eq:exhalation_fct}.
One important conclusion we draw from the observations in this section and the previous section is that the average flow velocity is not a sufficient indicator for the assessment of the filtration efficiency of an imperfectly fitted face mask if the actual air flow pattern exhibits dynamics beyond constant.

\subsection{Impact of IP's Face Mask on IIA Statistics}
\begin{figure}[!t]
    \centering
    \includegraphics[width=.42\textwidth]{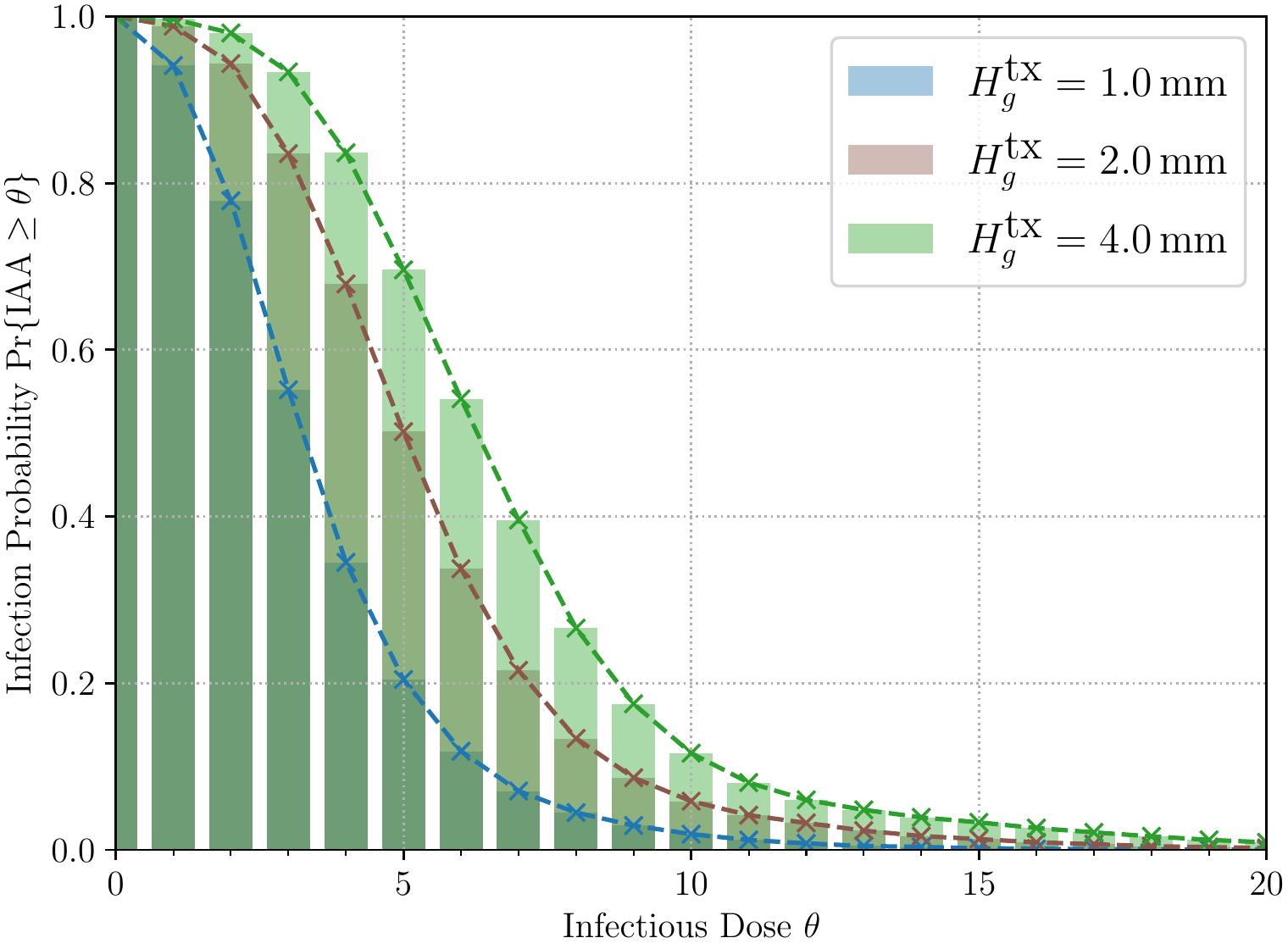}
    \caption{\small Cumulative empirical distribution of number of inhaled infectious aerosols IAA for different gap heights of \ac{IP}'s face mask.}
    \label{fig:gap_height_cdf}
\end{figure}

\vspace*{-0.5ex}
Finally, we examined the impact of the fit of the face mask of the \ac{IP} on the number of inhaled infectious aerosols $\IIA$ as defined in \eqref{eq:def_iia}.
From the cumulative empirical distribution of the \ac{RV} $\IIA$ shown in Fig.~\ref{fig:gap_height_cdf}, we observe that the infection probability, as defined in \eqref{eq:infection_probability}, is highly sensitive towards changes in $\Htx_g$ for infectious doses $\theta$ ranging from $2$ to $12$ particles.
However, for larger infectious doses, i.e., $\theta \gg 12$, the impact of $\Htx_g$ on the infection probability is rather small.
This is an interesting observation as it shows the power of the proposed model for identifying the regime in which the infection risk is sensitive to the fit of the \ac{IP}'s face mask.

%

%% file: conclusions.tex
\vspace*{-0.5ex}
In this paper, we presented a novel model for the impact of imperfectly fitted face masks on airborne virus transmission based on the \ac{MC} paradigm.
By formulating the problem of virus transmission as a communication problem, we obtained a statistical virus transmission model which reveals the impact of several physical face mask parameters on the virus transmission.
The proposed model is simple but extends existing models by encompassing the randomness of the emission and the inhalation of \acp{IA} due to the imperfectly fitted face masks of the \ac{IP} and the \ac{HP}, respectively.
With the help of our model, we could demonstrate the importance of the air flow dynamics for the assessment of the filtration efficiency of face masks.
This observation is especially relevant as virus transmission prevention concerns foremost random (transmission) events with low probability, i.e., statistical outliers, and, hence, risk assessment beyond first-order statistics is required.
Finally, we have demonstrated how the proposed model can be used to characterize the end-to-end transmission of \acp{IA} from an \ac{IP} to an \ac{HP} statistically when both individuals wear imperfectly fitted face masks.
In the broader context of \ac{MC}, our model provides a step towards understanding how the stochastic variability of the signaling molecules, which are commonly assumed to be identical, may impact the statistics of the end-to-end communication channel.

Possible extensions of the work presented in this paper include a more comprehensive modeling of the physical environment in which the virus transmission takes place and the validation of the proposed model with empirical data.

%% file: airborne_virus_spread_statistical_transmission_model.bbl

\begin{thebibliography}{26}


\ifx \showCODEN    \undefined \def \showCODEN     #1{\unskip}     \fi
\ifx \showDOI      \undefined \def \showDOI       #1{#1}\fi
\ifx \showISBNx    \undefined \def \showISBNx     #1{\unskip}     \fi
\ifx \showISBNxiii \undefined \def \showISBNxiii  #1{\unskip}     \fi
\ifx \showISSN     \undefined \def \showISSN      #1{\unskip}     \fi
\ifx \showLCCN     \undefined \def \showLCCN      #1{\unskip}     \fi
\ifx \shownote     \undefined \def \shownote      #1{#1}          \fi
\ifx \showarticletitle \undefined \def \showarticletitle #1{#1}   \fi
\ifx \showURL      \undefined \def \showURL       {\relax}        \fi
\providecommand\bibfield[2]{#2}
\providecommand\bibinfo[2]{#2}
\providecommand\natexlab[1]{#1}
\providecommand\showeprint[2][]{arXiv:#2}

\bibitem[\protect\citeauthoryear{Azimi et~al\mbox{.}}{Azimi
  et~al\mbox{.}}{2021}]%
        {azimi21}
\bibfield{author}{\bibinfo{person}{Parham Azimi} {et~al\mbox{.}}}
  \bibinfo{year}{2021}\natexlab{}.
\newblock \showarticletitle{Mechanistic transmission modeling of {COVID}-19 on
  the Diamond Princess cruise ship demonstrates the importance of aerosol
  transmission}.
\newblock \bibinfo{journal}{\emph{Proc. Natl. Acad. Sci. U.S.A}}
  \bibinfo{volume}{118}, \bibinfo{number}{8} (\bibinfo{date}{Feb.}
  \bibinfo{year}{2021}).
\newblock
\showISSN{0027-8424}
\urldef\tempurl%
\url{https://doi.org/10.1073/pnas.2015482118}
\showDOI{\tempurl}


\bibitem[\protect\citeauthoryear{Barros et~al\mbox{.}}{Barros
  et~al\mbox{.}}{2020}]%
        {barros20}
\bibfield{author}{\bibinfo{person}{Michael~Taynnan Barros} {et~al\mbox{.}}}
  \bibinfo{year}{2020}\natexlab{}.
\newblock \bibinfo{title}{Molecular Communications in Viral Infections
  Research: Modelling, Experimental Data and Future Directions}.
\newblock
\newblock
\showeprint[arxiv]{2011.00002}~[q-bio.OT]
\urldef\tempurl%
\url{https://arxiv.org/abs/2011.00002}
\showURL{%
\tempurl}


\bibitem[\protect\citeauthoryear{Bhagat, Davies~Wykes, Dalziel, and
  Linden}{Bhagat et~al\mbox{.}}{2020}]%
        {bhagat20}
\bibfield{author}{\bibinfo{person}{Rajesh~K. Bhagat}, \bibinfo{person}{M.~S.
  Davies~Wykes}, \bibinfo{person}{Stuart~B. Dalziel}, {and}
  \bibinfo{person}{P.~F. Linden}.} \bibinfo{year}{2020}\natexlab{}.
\newblock \showarticletitle{Effects of ventilation on the indoor spread of
  {COVID}-19}.
\newblock \bibinfo{journal}{\emph{J. Fluid Mech.}}  \bibinfo{volume}{903}
  (\bibinfo{date}{Nov.} \bibinfo{year}{2020}), \bibinfo{pages}{F1}.
\newblock
\urldef\tempurl%
\url{https://doi.org/10.1017/jfm.2020.720}
\showDOI{\tempurl}


\bibitem[\protect\citeauthoryear{{Chahibi} and {Akyildiz}}{{Chahibi} and
  {Akyildiz}}{2014}]%
        {chahibi14}
\bibfield{author}{\bibinfo{person}{Y. {Chahibi}} {and} \bibinfo{person}{I.~F.
  {Akyildiz}}.} \bibinfo{year}{2014}\natexlab{}.
\newblock \showarticletitle{Molecular Communication Noise and Capacity Analysis
  for Particulate Drug Delivery Systems}.
\newblock \bibinfo{journal}{\emph{IEEE Trans. Commun.}} \bibinfo{volume}{62},
  \bibinfo{number}{11} (\bibinfo{date}{Nov.} \bibinfo{year}{2014}),
  \bibinfo{pages}{3891--3903}.
\newblock


\bibitem[\protect\citeauthoryear{Cheng et~al\mbox{.}}{Cheng
  et~al\mbox{.}}{2020}]%
        {cheng20}
\bibfield{author}{\bibinfo{person}{Yafang Cheng} {et~al\mbox{.}}}
  \bibinfo{year}{2020}\natexlab{}.
\newblock \showarticletitle{Distinct regimes of particle and virus abundance
  explain face mask efficacy for {COVID}-19}.
\newblock \bibinfo{journal}{\emph{medRxiv}} (\bibinfo{year}{2020}).
\newblock
\urldef\tempurl%
\url{https://doi.org/10.1101/2020.09.10.20190348}
\showDOI{\tempurl}


\bibitem[\protect\citeauthoryear{Devroye}{Devroye}{1986}]%
        {devroye86}
\bibfield{author}{\bibinfo{person}{Luc Devroye}.}
  \bibinfo{year}{1986}\natexlab{}.
\newblock \showarticletitle{General principles in random variate generation}.
\newblock In \bibinfo{booktitle}{\emph{Non-uniform Random Variate Generation}
  (\bibinfo{edition}{1} ed.)}. \bibinfo{publisher}{Springer},
  \bibinfo{pages}{27--82}.
\newblock


\bibitem[\protect\citeauthoryear{Fatih and Baris}{Fatih and Baris}{2021}]%
        {gulec21}
\bibfield{author}{\bibinfo{person}{Gulec Fatih} {and} \bibinfo{person}{Atakan
  Baris}.} \bibinfo{year}{2021}\natexlab{}.
\newblock \showarticletitle{Fluid dynamics-based distance estimation algorithm
  for macroscale molecular communication}.
\newblock \bibinfo{journal}{\emph{Nano Commun. Netw.}}  \bibinfo{volume}{28}
  (\bibinfo{date}{Feb.} \bibinfo{year}{2021}), \bibinfo{pages}{100351}.
\newblock
\showISSN{1878-7789}
\urldef\tempurl%
\url{https://doi.org/10.1016/j.nancom.2021.100351}
\showDOI{\tempurl}


\bibitem[\protect\citeauthoryear{Gandhi and Marr}{Gandhi and Marr}{2021}]%
        {gandhi21}
\bibfield{author}{\bibinfo{person}{Monica Gandhi} {and}
  \bibinfo{person}{Linsey~C. Marr}.} \bibinfo{year}{2021}\natexlab{}.
\newblock \showarticletitle{Uniting Infectious Disease and Physical Science
  Principles on the Importance of Face Masks for {COVID}-19}.
\newblock \bibinfo{journal}{\emph{Med}} \bibinfo{volume}{2},
  \bibinfo{number}{1} (\bibinfo{date}{Jan.} \bibinfo{year}{2021}),
  \bibinfo{pages}{29--32}.
\newblock
\showISSN{2666-6340}
\urldef\tempurl%
\url{https://doi.org/10.1016/j.medj.2020.12.008}
\showDOI{\tempurl}


\bibitem[\protect\citeauthoryear{Gulec and Atakan}{Gulec and Atakan}{2020}]%
        {gulec20}
\bibfield{author}{\bibinfo{person}{Fatih Gulec} {and} \bibinfo{person}{Baris
  Atakan}.} \bibinfo{year}{2020}\natexlab{}.
\newblock \showarticletitle{A molecular communication perspective on airborne
  pathogen transmission and reception via droplets generated by coughing and
  sneezing}.
\newblock \bibinfo{journal}{\emph{arXiv preprint arXiv:2007.07598}}
  (\bibinfo{year}{2020}).
\newblock


\bibitem[\protect\citeauthoryear{Howard et~al\mbox{.}}{Howard
  et~al\mbox{.}}{2021}]%
        {howard21}
\bibfield{author}{\bibinfo{person}{Jeremy Howard} {et~al\mbox{.}}}
  \bibinfo{year}{2021}\natexlab{}.
\newblock \showarticletitle{An evidence review of face masks against
  {COVID-19}}.
\newblock \bibinfo{journal}{\emph{Proc. Natl. Acad. Sci. U.S.A}}
  \bibinfo{volume}{118}, \bibinfo{number}{4} (\bibinfo{date}{Jan.}
  \bibinfo{year}{2021}).
\newblock
\showISSN{0027-8424}
\urldef\tempurl%
\url{https://doi.org/10.1073/pnas.2014564118}
\showDOI{\tempurl}


\bibitem[\protect\citeauthoryear{Hu, Guo, Zhou, and Shi}{Hu
  et~al\mbox{.}}{2021}]%
        {hu21}
\bibfield{author}{\bibinfo{person}{Ben Hu}, \bibinfo{person}{Hua Guo},
  \bibinfo{person}{Peng Zhou}, {and} \bibinfo{person}{Zheng-Li Shi}.}
  \bibinfo{year}{2021}\natexlab{}.
\newblock \showarticletitle{Characteristics of {SARS-CoV-2} and {COVID-19}}.
\newblock \bibinfo{journal}{\emph{Nat. Rev. Microbiol.}} \bibinfo{volume}{19},
  \bibinfo{number}{3} (\bibinfo{date}{Mar.} \bibinfo{year}{2021}),
  \bibinfo{pages}{141--154}.
\newblock
\showISSN{1740-1534}
\urldef\tempurl%
\url{https://doi.org/10.1038/s41579-020-00459-7}
\showDOI{\tempurl}


\bibitem[\protect\citeauthoryear{{Khalid}, {Amin}, {Ahmed}, {Shihada}, and
  {Alouini}}{{Khalid} et~al\mbox{.}}{2019}]%
        {khalid19}
\bibfield{author}{\bibinfo{person}{M. {Khalid}}, \bibinfo{person}{O. {Amin}},
  \bibinfo{person}{S. {Ahmed}}, \bibinfo{person}{B. {Shihada}}, {and}
  \bibinfo{person}{M. {Alouini}}.} \bibinfo{year}{2019}\natexlab{}.
\newblock \showarticletitle{Communication through Breath: Aerosol
  Transmission}.
\newblock \bibinfo{journal}{\emph{IEEE Commun. Mag.}} \bibinfo{volume}{57},
  \bibinfo{number}{2} (\bibinfo{date}{Feb.} \bibinfo{year}{2019}),
  \bibinfo{pages}{33--39}.
\newblock
\urldef\tempurl%
\url{https://doi.org/10.1109/MCOM.2018.1800530}
\showDOI{\tempurl}


\bibitem[\protect\citeauthoryear{{Khalid}, {Amin}, {Ahmed}, {Shihada}, and
  {Alouini}}{{Khalid} et~al\mbox{.}}{2020}]%
        {khalid20}
\bibfield{author}{\bibinfo{person}{M. {Khalid}}, \bibinfo{person}{O. {Amin}},
  \bibinfo{person}{S. {Ahmed}}, \bibinfo{person}{B. {Shihada}}, {and}
  \bibinfo{person}{M.~S. {Alouini}}.} \bibinfo{year}{2020}\natexlab{}.
\newblock \showarticletitle{Modeling of Viral Aerosol Transmission and
  Detection}.
\newblock \bibinfo{journal}{\emph{IEEE Trans. Commun.}} \bibinfo{volume}{68},
  \bibinfo{number}{8} (\bibinfo{date}{Aug.} \bibinfo{year}{2020}),
  \bibinfo{pages}{4859--4873}.
\newblock
\urldef\tempurl%
\url{https://doi.org/10.1109/TCOMM.2020.2994191}
\showDOI{\tempurl}


\bibitem[\protect\citeauthoryear{Koca, Civas, Sahin, Ergonul, and Akan}{Koca
  et~al\mbox{.}}{2021}]%
        {koca21}
\bibfield{author}{\bibinfo{person}{Caglar Koca}, \bibinfo{person}{Meltem
  Civas}, \bibinfo{person}{Selin~M. Sahin}, \bibinfo{person}{Onder Ergonul},
  {and} \bibinfo{person}{Ozgur~B. Akan}.} \bibinfo{year}{2021}\natexlab{}.
\newblock \showarticletitle{Molecular Communication Theoretical Modeling and
  Analysis of {SARS-CoV2} Transmission in Human Respiratory System}.
\newblock  (\bibinfo{year}{2021}).
\newblock
\showeprint[arxiv]{2011.05154}~[physics.bio-ph]
\urldef\tempurl%
\url{https://arxiv.org/abs/2011.05154}
\showURL{%
\tempurl}


\bibitem[\protect\citeauthoryear{Lei, Yang, Zhuang, and Roberge}{Lei
  et~al\mbox{.}}{2012}]%
        {lei12}
\bibfield{author}{\bibinfo{person}{Zhipeng Lei}, \bibinfo{person}{James Yang},
  \bibinfo{person}{Ziqing Zhuang}, {and} \bibinfo{person}{Raymond Roberge}.}
  \bibinfo{year}{2012}\natexlab{}.
\newblock \showarticletitle{Simulation and Evaluation of Respirator Faceseal
  Leaks Using Computational Fluid Dynamics and Infrared Imaging}.
\newblock \bibinfo{journal}{\emph{Ann. Occup. Hyg.}} \bibinfo{volume}{57},
  \bibinfo{number}{4} (\bibinfo{date}{12} \bibinfo{year}{2012}),
  \bibinfo{pages}{493--506}.
\newblock
\showISSN{0003-4878}
\urldef\tempurl%
\url{https://doi.org/10.1093/annhyg/mes085}
\showDOI{\tempurl}


\bibitem[\protect\citeauthoryear{Lelieveld et~al\mbox{.}}{Lelieveld
  et~al\mbox{.}}{2020}]%
        {lelieveld20}
\bibfield{author}{\bibinfo{person}{Jos Lelieveld} {et~al\mbox{.}}}
  \bibinfo{year}{2020}\natexlab{}.
\newblock \showarticletitle{Model Calculations of Aerosol Transmission and
  Infection Risk of {COVID}-19 in Indoor Environments}.
\newblock \bibinfo{journal}{\emph{Int. J. Environ. Res. Public Health}}
  \bibinfo{volume}{17}, \bibinfo{number}{21} (\bibinfo{date}{Nov.}
  \bibinfo{year}{2020}).
\newblock
\showISSN{1660-4601}
\urldef\tempurl%
\url{https://doi.org/10.3390/ijerph17218114}
\showDOI{\tempurl}


\bibitem[\protect\citeauthoryear{Mittal, Meneveau, and Wu}{Mittal
  et~al\mbox{.}}{2020a}]%
        {mittal20a}
\bibfield{author}{\bibinfo{person}{Rajat Mittal}, \bibinfo{person}{Charles
  Meneveau}, {and} \bibinfo{person}{Wen Wu}.} \bibinfo{year}{2020}\natexlab{a}.
\newblock \showarticletitle{A mathematical framework for estimating risk of
  airborne transmission of {COVID-19} with application to face mask use and
  social distancing}.
\newblock \bibinfo{journal}{\emph{Phys. Fluids}} \bibinfo{volume}{32},
  \bibinfo{number}{10} (\bibinfo{date}{Oct.} \bibinfo{year}{2020}),
  \bibinfo{pages}{101903}.
\newblock
\urldef\tempurl%
\url{https://doi.org/10.1063/5.0025476}
\showDOI{\tempurl}
\showeprint{https://doi.org/10.1063/5.0025476}


\bibitem[\protect\citeauthoryear{Mittal, Ni, and Seo}{Mittal
  et~al\mbox{.}}{2020b}]%
        {mittal20}
\bibfield{author}{\bibinfo{person}{Rajat Mittal}, \bibinfo{person}{Rui Ni},
  {and} \bibinfo{person}{Jung-Hee Seo}.} \bibinfo{year}{2020}\natexlab{b}.
\newblock \showarticletitle{The flow physics of {COVID}-19}.
\newblock \bibinfo{journal}{\emph{J. Fluid Mech.}}  \bibinfo{volume}{894}
  (\bibinfo{date}{Mai} \bibinfo{year}{2020}), \bibinfo{pages}{F2}.
\newblock
\urldef\tempurl%
\url{https://doi.org/10.1017/jfm.2020.330}
\showDOI{\tempurl}


\bibitem[\protect\citeauthoryear{Morawska and Milton}{Morawska and
  Milton}{2020}]%
        {morawska20}
\bibfield{author}{\bibinfo{person}{Lidia Morawska} {and}
  \bibinfo{person}{Donald~K Milton}.} \bibinfo{year}{2020}\natexlab{}.
\newblock \showarticletitle{It Is Time to Address Airborne Transmission of
  Coronavirus Disease 2019 ({COVID}-19)}.
\newblock \bibinfo{journal}{\emph{Clin. Infect. Dis.}} \bibinfo{volume}{71},
  \bibinfo{number}{9} (\bibinfo{date}{Nov.} \bibinfo{year}{2020}),
  \bibinfo{pages}{2311--2313}.
\newblock
\showISSN{1058-4838}
\urldef\tempurl%
\url{https://doi.org/10.1093/cid/ciaa939}
\showDOI{\tempurl}


\bibitem[\protect\citeauthoryear{Nakano, Eckford, and Haraguchi}{Nakano
  et~al\mbox{.}}{2013}]%
        {nakano13}
\bibfield{author}{\bibinfo{person}{Tadashi Nakano}, \bibinfo{person}{Andrew~W.
  Eckford}, {and} \bibinfo{person}{Tokuko Haraguchi}.}
  \bibinfo{year}{2013}\natexlab{}.
\newblock \bibinfo{booktitle}{\emph{Molecular Communication}
  (\bibinfo{edition}{1} ed.)}.
\newblock \bibinfo{publisher}{Cambridge University Press}.
\newblock
\urldef\tempurl%
\url{https://doi.org/10.1017/CBO9781139149693}
\showDOI{\tempurl}


\bibitem[\protect\citeauthoryear{Peri{\'c} and Peri{\'c}}{Peri{\'c} and
  Peri{\'c}}{2020}]%
        {peric20}
\bibfield{author}{\bibinfo{person}{Robinson Peri{\'c}} {and}
  \bibinfo{person}{Milovan Peri{\'c}}.} \bibinfo{year}{2020}\natexlab{}.
\newblock \showarticletitle{Analytical and numerical investigation of the
  airflow in face masks used for protection against {COVID}-19
  virus--implications for mask design and usage}.
\newblock \bibinfo{journal}{\emph{J. Appl. Fluid Mech.}} \bibinfo{volume}{13},
  \bibinfo{number}{6} (\bibinfo{date}{Nov} \bibinfo{year}{2020}),
  \bibinfo{pages}{1911--1923}.
\newblock
\showISSN{17353572}


\bibitem[\protect\citeauthoryear{Riediker and Tsai}{Riediker and Tsai}{2020}]%
        {riediker20}
\bibfield{author}{\bibinfo{person}{Michael Riediker} {and}
  \bibinfo{person}{Dai-Hua Tsai}.} \bibinfo{year}{2020}\natexlab{}.
\newblock \showarticletitle{Estimation of Viral Aerosol Emissions From
  Simulated Individuals With Asymptomatic to Moderate Coronavirus Disease
  2019}.
\newblock \bibinfo{journal}{\emph{JAMA Netw. Open}} \bibinfo{volume}{3},
  \bibinfo{number}{7} (\bibinfo{date}{Jul.} \bibinfo{year}{2020}),
  \bibinfo{pages}{e2013807--e2013807}.
\newblock
\showISSN{2574-3805}
\urldef\tempurl%
\url{https://doi.org/10.1001/jamanetworkopen.2020.13807}
\showDOI{\tempurl}


\bibitem[\protect\citeauthoryear{Russo and Khalifa}{Russo and Khalifa}{2011}]%
        {russo11}
\bibfield{author}{\bibinfo{person}{Jackie~Sue Russo} {and}
  \bibinfo{person}{Ezzat Khalifa}.} \bibinfo{year}{2011}\natexlab{}.
\newblock \showarticletitle{Computational study of breathing methods for
  inhalation exposure}.
\newblock \bibinfo{journal}{\emph{HVAC\&R Res.}} \bibinfo{volume}{17},
  \bibinfo{number}{4} (\bibinfo{date}{Aug.} \bibinfo{year}{2011}),
  \bibinfo{pages}{419--431}.
\newblock
\urldef\tempurl%
\url{https://doi.org/10.1080/10789669.2011.578701}
\showDOI{\tempurl}
\showeprint{https://www.tandfonline.com/doi/pdf/10.1080/10789669.2011.578701}


\bibitem[\protect\citeauthoryear{Schurwanz, Hoeher, Bhattacharjee, Damrath,
  Stratmann, and Dressler}{Schurwanz et~al\mbox{.}}{2020}]%
        {schurwanz20}
\bibfield{author}{\bibinfo{person}{Max Schurwanz}, \bibinfo{person}{Peter~Adam
  Hoeher}, \bibinfo{person}{Sunasheer Bhattacharjee}, \bibinfo{person}{Martin
  Damrath}, \bibinfo{person}{Lukas Stratmann}, {and} \bibinfo{person}{Falko
  Dressler}.} \bibinfo{year}{2020}\natexlab{}.
\newblock \showarticletitle{Duality between coronavirus transmission and
  air-based macroscopic molecular communication}.
\newblock  (\bibinfo{year}{2020}).
\newblock
\urldef\tempurl%
\url{https://arxiv.org/abs/2009.04966}
\showURL{%
\tempurl}


\bibitem[\protect\citeauthoryear{{Viola} et~al\mbox{.}}{{Viola}
  et~al\mbox{.}}{2021}]%
        {viola21}
\bibfield{author}{\bibinfo{person}{I.~M. {Viola}} {et~al\mbox{.}}}
  \bibinfo{year}{2021}\natexlab{}.
\newblock \showarticletitle{Face Coverings, Aerosol Dispersion and Mitigation
  of Virus Transmission Risk}.
\newblock \bibinfo{journal}{\emph{IEEE Open J. Eng. Med. Biol.}}
  \bibinfo{volume}{2} (\bibinfo{date}{Jan.} \bibinfo{year}{2021}),
  \bibinfo{pages}{26--35}.
\newblock
\urldef\tempurl%
\url{https://doi.org/10.1109/OJEMB.2021.3053215}
\showDOI{\tempurl}


\bibitem[\protect\citeauthoryear{Wiersinga, Rhodes, Cheng, Peacock, and
  Prescott}{Wiersinga et~al\mbox{.}}{2020}]%
        {wiersinga20}
\bibfield{author}{\bibinfo{person}{W.~Joost Wiersinga}, \bibinfo{person}{Andrew
  Rhodes}, \bibinfo{person}{Allen~C. Cheng}, \bibinfo{person}{Sharon~J.
  Peacock}, {and} \bibinfo{person}{Hallie~C. Prescott}.}
  \bibinfo{year}{2020}\natexlab{}.
\newblock \showarticletitle{Pathophysiology, Transmission, Diagnosis, and
  Treatment of Coronavirus Disease 2019 ({COVID}-19): A Review}.
\newblock \bibinfo{journal}{\emph{JAMA}} \bibinfo{volume}{324},
  \bibinfo{number}{8} (\bibinfo{date}{Aug.} \bibinfo{year}{2020}),
  \bibinfo{pages}{782--793}.
\newblock
\showISSN{0098-7484}
\urldef\tempurl%
\url{https://doi.org/10.1001/jama.2020.12839}
\showDOI{\tempurl}


\end{thebibliography}
